\documentclass[reprint,superscriptaddress,nobibnotes,amsmath,amssymb,twocolumn,prb]{revtex4-2}
\usepackage[pdftex]{graphicx}
\usepackage[english]{babel}
\usepackage{physics}
\usepackage{tabularx}
\usepackage{float}
\usepackage{amssymb}   
\usepackage{amsmath}   
\usepackage{dsfont}   
\usepackage{epsfig}
\usepackage{dcolumn}   
\usepackage{bm}        
\usepackage{amstext}
\usepackage{epsfig}
\usepackage{mdframed}
\newcommand{\be}{\begin{equation}}
\newcommand{\ee}{\end{equation}}

\newcommand{\LCPQ}{Laboratoire de Chimie et Physique Quantiques (UMR 5626), CNRS and Universit\'e de Toulouse, France}

\begin{document}
\title{Thermodynamics of quantum oscillators}
\author{Michel Caffarel}
\email[]{caffarel@irsamc.ups-tlse.fr}
\affiliation{\LCPQ}
\begin{abstract}
In this work, we present a compact analytical approximation for the quantum partition function 
of systems composed of quantum oscillators. The proposed formula is general and applicable 
to an arbitrary number of oscillators 
described by a rather general class of potential energy functions (not necessarily polynomials).
Starting from the exact path integral expression of the partition function, 
we introduce a temperature-dependent Gaussian approximation for the high-temperature propagator and, 
then, invoke a principle of minimal sensitivity 
to minimize the error. This leads to a system of coupled nonlinear equations 
whose solution yields the optimal parameters of the Gaussian approximation. 
The resulting approximate partition function accurately reproduces 
thermodynamic quantities such as the free energy, average energy, and specific heat -even at zero temperature- 
with typical relative errors in the range of about 1--5\%. The accuracy deteriorates only moderately 
when the anharmonicity and coupling strengths are increased.
We illustrate the performance of our analytical formula with numerical results for systems of up 
to ten coupled anharmonic oscillators. 
These results are compared to "exact" numerical results 
obtained via Hamiltonian diagonalization for small systems and Path Integral Monte Carlo simulations for larger ones.

\end{abstract}
\noindent
\maketitle
\section{Introduction}
In this work, we address the problem of evaluating the quantum partition function and its derived thermodynamic properties, 
such as the mean energy and the specific heat\cite{LandauLifshitzSP}. This problem is known to be an important and challenging task in statistical physics, 
particularly for large systems.
The method presented here is designed for general systems of quantum oscillators. By “general,” we mean that it is applicable to an arbitrary 
number of oscillators and does not require specific restrictions on the form of the potential energy function 
(which is therefore not limited to polynomial forms, as is often assumed). Furthermore, the approach can, in principle, 
be applied to arbitrary anharmonicities and couplings between oscillators, although, as in most approaches, its accuracy decreases when the anharmonicity and coupling strengths become too large.

As an important example in chemical physics, we mention the problem of computing the thermodynamic properties of large molecules. 
Such calculations can provide insight into the stability of biomolecular structures (proteins, DNA, RNA) as well as into 
the relative stability of different conformations. For instance, in drug design, calculating free-energy differences 
allows one to evaluate the binding affinity between a drug molecule and its biological target,
a central quantity in structure-based drug discovery\cite{Gilson2007,Chipot2007,Mobley2017}.
More generally, many biological processes are governed by thermodynamic balances. Beyond these applications 
in chemical physics, quantum oscillators also play a central role in a wide range of other scientific domains. 
Examples include phonons in solids\cite{srivastava1990physics}, quantum field theory ({\it e.g.}, the quantization of the electromagnetic field\cite{boyd2020nonlinear} and
$\phi^4$ theory\cite{peskin1995introduction},\cite{ZinnJustin2021}), nonlinear optical phenomena\cite{boyd2020nonlinear},
quantum decoherence\cite{zurek2003decoherence}, and qubits in superconducting circuits\cite{kjaergaard2020superconducting}.

In practice, thermodynamic quantities are usually computed using numerical methods, most notably Monte Carlo or 
molecular dynamics simulations (this is the case either for the classical or quantum partition function, but under different 
theoretical frameworks). 
While these methods can provide reliable estimates of equilibrium properties, they typically require large-scale simulations and substantial computational resources, whose cost increases rapidly with the size of the system.

An alternative route to evaluate thermodynamic properties is to employ explicit analytical representations 
of the partition function; 
this is the focus of the present work. By using an explicit formula, the computational burden 
is drastically reduced, 
allowing much larger systems to be studied. Furthermore, an analytic expression makes it possible 
to directly follow the dependence 
of thermodynamic quantities on the parameters of the system. This can provide clearer physical insight, for instance by revealing 
limiting behaviors or mechanisms that may remain hidden in purely numerical approaches. 
However, the price to pay is the error introduced when approximating the exact unknown expression. 
To be of practical interest, an approximate formula for the partition function 
must faithfully represent the exact one 
while providing sufficiently accurate quantitative predictions for thermodynamic properties, 
in particular 
when compared with "exact" numerical results obtained from extensive simulations.

In the present work, we aim to derive an original analytical formula that meets these requirements. 
The important result of this work is that the applications presented here suggest that our formula could be a promising candidate. 
For the simplest case of a one-dimensional quartic oscillator (potential of the form 
$\frac{1}{2} \omega^2 x^2 + g x^4$, see {\it e.g.} [\onlinecite{TurbinerDelValle2023}]), we have already shown\cite{osc1} 
that our analytical formula reproduces, quite remarkably, both the {\it qualitative} and {\it quantitative} features of the exact 
partition function. In summary, we have obtained the following results:  i) the free energy 
is reproduced with an accuracy of a few percent over the entire range of temperatures 
and coupling strengths, ii) both the harmonic limit ($g \rightarrow 0$) 
and the classical high-temperature limit are recovered exactly, iii)
the divergence of the weak-coupling perturbation series for the ground-state energy, characterized by 
the factorial growth of perturbative coefficients, is also correctly reproduced (by far a nontrivial result, see {\it e.g.} [\onlinecite{Bender_1969}]). 
The same holds for the functional form of the strong-coupling expansion, with coefficients that are found to be reasonably accurate in both regimes, iv)
a large number of accurate excited-state energies can be extracted from the partition function\cite{osc2}.

In the present work, we derive the general formula for an arbitrary number of coupled anharmonic oscillators 
and investigate its accuracy by considering several test systems. We first study simple cases for which "exact" numerical results 
can be obtained through Hamiltonian diagonalization. We then examine larger systems involving up to ten coupled anharmonic oscillators, 
for which our results are compared with Path Integral Monte Carlo simulations. Overall, we find that the approximate partition function 
reproduces thermodynamic quantities, such as the free energy, average energy, and specific heat, even at zero temperature, 
with an accuracy comparable to that obtained in the one-dimensional case. 
In the absence of couplings between oscillators, the partition function factorizes into a product of one-dimensional partition functions 
corresponding to independent oscillators. The harmonic limit and the classical high-temperature limit are also recovered. 
The derivation of the excited-state spectrum is left for future work.

To derive our formula, we start from the exact path-integral representation of the quantum partition function introduced by Feynman\cite{feynman1972statistical}. This representation is known to be particularly convenient for the study of large systems. 
The next step is to introduce an approximation scheme for the path integral, which leads to an explicit and compact expression for 
the partition function. To the best of our knowledge, this approximation scheme has not been previously proposed in the literature.

Finally, it seems appropriate to present the final expression for the partition function already in this introduction. 
Displaying the formula here will give to the reader a general view
of its structure and level of complexity. The detailed derivation is provided in the main body of the paper.\\

The formula reads
\be
{\cal Z}(\beta)=  {\cal N}(\beta) \prod_{k=1}^N  \frac{1}{e^{\frac{\beta {\Omega}_{gk}(\beta) }{2}}
- e^{-\frac{\beta {\Omega}_{gk}(\beta)}{2}}}
\label{Z_intro}
\ee
where $N$ is the number of quantum oscillators, $\beta$ the inverse temperature ($\beta=\frac{1}{k_B T}$), $ {\cal N}(\beta) $ a prefactor, and ${\Omega}_{gk}(\beta)$ a set of 
$N$ temperature-dependent frequencies. The frequencies, ${\Omega}_{gk}$ are obtained from the eigenvalues of the matrix
\be
M=
\begin{bmatrix}
\omega^2_{g1} & K_{g12} & \cdots & K_{g1N} \\
K_{g12} & \omega^2_{g2} & \cdots & K_{g2N} \\
\vdots & \vdots & \ddots & \vdots \\
K_{g1N} & K_{g2N} & \cdots & \omega^2_{gN}
\label{matrixM}
\end{bmatrix}
\ee
where the quantities $\{\omega_{gk},K_{gkl}\}$ are some temperature-dependent (Gaussian) parameters obtained by fitting the exact short-time 
propagator involved in the path integral by a Gaussian propagator.
For all $k$ and pairs $(k,l)$ the Gaussian parameters are given by 
\be
\omega_{gk}= \sqrt{\frac{C^{-1}_{kk}}{\tau_k}}
\label{FIT1}
\ee
\be
K_{gkl}= \frac{C^{-1}_{kl}}{\tau_{kl}}.
\label{FIT2}
\ee
Here, $C$ is the covariance matrix defined as
$C_{kl}= \langle \qty(x_k - \langle x_k \rangle_\pi) \qty(x_l - \langle x_l \rangle_\pi) \rangle_\pi$. For a potential 
including only one-body terms and pairwise couplings between vibrational modes, 
$\pi$ is the following "multi time-step" distribution
\be
\pi= \frac{e^{ -\sum_{k=1}^N \tau_k V_k(x_k) - \sum_{\substack{1 \leq k < l \leq N}} \tau_{kl} V_{kl}(x_k,x_l)}}
{ \int d{\bf x} e^{ -\sum_{k=1}^N \tau_k V_k(x_k) - \sum_{\substack{1 \leq k < l \leq N}} \tau_{kl} V_{kl}}}.
\label{pi_multiple}
\ee
The quantities $\tau_k$ and $\tau_{kl}$ are generalized time-steps associated with the Gaussian parameters $\omega_{gk}$ and $K_{gkl}$, respectively.
They are obtained by solving a set of 
implicit non-linear coupled equations given by
\be
\tau_k = \frac{ 2
        \sum_{i=1}^N \frac{\partial {\Omega}_{gi}} {\partial  \omega_{gk}} \frac{1}{ \Omega_{gi}}
}
{ \sum_{i=1}^N  \frac{\partial {\Omega}_{gi}} {\partial \omega_{gk}}
\coth{ \frac{ \beta  \Omega_{gi}}{2} }}.
\label{solv1}
\ee
and
\be
\tau_{kl} =\frac{ 2
        \sum_{i=1}^N \frac{\partial {\Omega}_{gi}} {\partial K_{gkl}} \frac{1}{ \Omega_{gi}}
}
{ \sum_{i=1}^N  \frac{\partial {\Omega}_{gi}} {\partial K_{gkl}}
\coth{ \frac{ \beta  \Omega_{gi}}{2} }}.
\label{solv_fin}
\ee
Some freedom of choice exists for the prefactor ${\cal N}(\beta)$. As we shall show, a natural and accurate choice is
\be
{\cal N}(\beta) = 
\prod_{k=1}^N \qty(\frac{I_k}{I_{Gk}})^{\frac{\beta}{\tau_k}}  e^{ \frac{\beta}{2} (b_g, M^{-1} b_g)}
\label{Nn0}
\ee
where $I_k$ and $I_{Gk}$ are one-dimensional integrals given by
\be
I_k= \int dx_k e^{-\tau_k V_k(x_k)}
\label{Ik0}
\ee
and
\be
I_{Gk}= \int dx_k e^{-\tau_k \qty( \frac{1}{2} \omega^2_{gk} x^2_k + b_{gk} x_k)}.
\label{IGk0}
\ee
The quantities $b_{gk}$
appearing in Eqs.(\ref{Nn0}),(\ref{Ik0}) and (\ref{IGk0}) are additional Gaussian parameters resulting from the fit, in the same way as
the parameters $\omega_{gk}$ and $K_{gkl}$.
They are given by
\be
b_{gk}=-\frac{\qty( C^{-1} \mu)_k}{\tau_k}.
\label{bk}
\ee
where $\mu$ is the vector of first moments of $\pi$, $\mu_k= \langle x_k \rangle_\pi$.\\

As seen from the various expressions above, the dependence of the partition function on the temperature turns out to be rather complex.
Note also that the formula contains no adjustable parameters.
\\

The organization of the paper is as follows. In Section~\ref{theory}, we present the full derivation of our partition function in detail. 
Section~\ref{applications} is devoted to numerical applications. 
We begin with small systems ($N=1,2$, and 3 oscillators) for which the exact partition function can be known numerically with high accuracy. 
We then move on to a system of coupled quartic oscillators of increasing size, up to $N=10$. 
The results obtained with our model partition function are compared to those from path integral Monte Carlo (PIMC) simulations. 
Finally, in Section~\ref{conclusion}, we present some concluding remarks and perspectives.

\section{Theory}
\label{theory}
In this work, we consider the following Hamiltonian for a system of $N$ quantum oscillators
\begin{equation}
H = -\frac{1}{2} \sum_{k=1}^N \frac{\partial^2}{\partial x_k^2} + V(\mathbf{x}),
\label{H}
\end{equation}
where $x_k$ denotes the displacement of the $k$-th oscillator from equilibrium, $V(\mathbf{x})$ is the total potential energy,
and $\mathbf{x} = (x_1, \ldots, x_N)$ the vector of displacements.
Note that the coordinates $x_k$ may have been previously rescaled to absorb the oscillator masses, making all masses equal to one 
without loss of generality.
In this work, we will consider potential functions that involve only pairwise couplings between oscillators. 
More precisely, we consider potentials of the form
\be
V(\mathbf{x})= \sum_{k=1}^N V_k(x_k) + \sum_{\substack{1 \leq k < l \leq N}} V_{kl}(x_k,x_l).
\label{V}
\ee
In fact, much of the theory remains valid for more general coupling schemes and could be extended.
However, to avoid unnecessary complications in the exposition, we postpone the presentation of such extensions to future work. No specific constraints are imposed on the analytical form of $V_k$ and $V_{kl}$.
However, we require $V$ to be bounded from below and 
$V \rightarrow +\infty$ as $|{\bf x}| \rightarrow \infty$. These conditions are standard for 
generic oscillators; they lead to a discrete energy spectrum. 
However, they exclude cases where oscillations can be unbounded, such as for the Morse 
oscillator\cite{morse1929diatomic,landau1977quantum}.
In the illustrative simulations to follow, we will use a standard polynomial form of order 4 for the potential, namely $V=\sum_{k,l} c_{kl} x^m_k x^n_l$ 
with $m+n \le 4$, but we insist that non-polynomial forms can also be considered.

In order to derive our fundamental formula, we start from 
the exact path integral representation of the partition function\cite{feynman1972statistical},
\be
Z(\beta) = \lim_{n \rightarrow \infty} Z_n(\beta)
\label{Z}
\ee
with
$$
Z_n(\beta)=
\qty( \frac{1}{\sqrt{ 2 \pi \tau}})^{nN} \int \prod_{i=1}^n \prod_{k=1}^N dx^i_k
$$
\be
e^{-\frac{1}{\tau} \sum_{i,j}^n \sum_{k=1}^N x^i_k A_{ij} x^j_k -\tau \sum_{i=1}^n V({\bf x}^i)}
\label{Z_path}
\ee
Here, $\beta$ is the inverse temperature, $n$ the number of beads,
$\tau$ the time-step equal to $\tau=\frac{\beta}{n}$, $x^i_k$ the displacement of the $k$-th oscillator at time $t_i=i\tau$, and
${\bf x}^i=(x^i_1,...,x^i_N)$.
$A$ is the matrix associated with the kinetic part; it is given by
\be
A_{ij}= \delta_{ij}-\frac{1}{2} \qty(\delta_{i+1\;j} + \delta_{i-1 \;j})
\label{matrixA}
\ee
with $A_{1n}=A_{n1}=-\frac{1}{2}$ (periodic boundary conditions in time).

The central approximation of this work consists in replacing, at each time slice of the path integral, the exact potential contribution
$e^{-\tau V}$
by the corresponding contribution obtained from a quadratic approximation of the potential.
Precisely, the replacement is made for the normalized contributions as follows
\be
\frac{e^{-\tau V({\bf x})}}{\int d {\bf x} e^{-\tau V({\bf x})}} \rightarrow \frac{e^{-\tau V_G({\bf x},\tau)}}{\int d {\bf x}
e^{-\tau V_G({\bf x},\tau)}}.
\label{replaceV}
\ee
The quadratic potential is chosen under the following most general form
\be
V_G({\bf x},\tau)= \frac{1}{2} \sum_{k=1}^N \omega_{gk}^2(\tau) x^2_k +  b_{gk}(\tau) x_k + \sum_{k < l}  K_{gkl}(\tau) x_k x_l,
\label{VG}
\ee
where $K$ is a matrix chosen symmetric without loss of generality.
Here, the quantities $\qty[\omega_{gk}(\tau),b_{gk}(\tau)]$ for $k=1$ to $N$ and $K_{gkl}(\tau)$ for $k < l$
play the role of "fitting" parameters, hereafter called Gaussian parameters. 
The number of
Gaussian parameters in $V_G$ is $M = 2 N + \frac{N (N-1)}{2}$.
These parameters will be chosen so as to minimize the approximation error.
At this point, it is most important to emphasize that the quadratic potential {\it depends explicitly on the time-step} $\tau$ 
via the Gaussian parameters.
This must be sharply contrasted with what is done in more conventional approaches -such as
mean-field-like methods- where the Hamiltonian and the action are made quadratic from the start, with parameters {\it independent} of $\tau$.

Another point worth mentioning is that, although the potential $V_G$ [Eq.~(\ref{VG})] represents the most general
form of a multidimensional quadratic potential, it can only provide a faithful description
of single-well potentials but not multi-well potentials.
In principle, this does not constitute a limitation of the present framework,
since only a global Gaussian approximation is required.
However, in practice, this approximation being not well suited to multimodal distributions, its accuracy may deteriorate, particularly at low temperatures. An accurate treatment of such systems would likely require a multi-Gaussian extension of the present approach. Although such an extension appears feasible, its development is beyond the scope of the present work.

To determine the optimal $\tau$-dependent Gaussian parameters
some freedom is available.
Here, we propose to set the parameter values by imposing the equality of the first and
second moments of the exact and Gaussian distributions involved in Eq.(\ref{replaceV}).
\be
\langle x_k \rangle_\pi = \langle x_k \rangle_{\pi_G}   \;\;\; k=1 \; {\rm to} \; N
\label{moment1}
\ee
and 
\be
\langle x_k x_l \rangle_\pi = \langle x_k x_l \rangle_{\pi_G}  \;\;\; k < l \; k,l=1 {\rm to} \; N,
\label{moment2}
\ee
where the distributions write 
\be
\pi = \frac{ e^{-\tau V} }{ I(\tau)}
\ee
and
\be
\pi_G = \frac{ e^{-\tau V_G} }{I_G(\tau)}.
\ee
Here, $I(\tau)$ and $I_G(\tau)$ denote the normalization of the exact and  Gaussian distributions, respectively.\\

Equations (\ref{moment1}, \ref{moment2}) constitute a system of $M$ equations with $M$ unknowns, where $M$ 
is the number of Gaussian parameters to be determined.
To solve this system, we begin by expressing the Gaussian distribution $\pi_G$ in the following form:
\be
\pi_G = \frac{e^{-\frac{1}{2} \qty( {\bf x}, A {\bf x}) - B {\bf x}}}{ \int d{\bf x} e^{- \frac{1}{2} \qty( {\bf x}, A {\bf x}) - B {\bf x}} }
\ee
where the matrix and vector components are defined as
$A_{kl}= \tau \qty( \omega^2_{gk} \delta_{kl} + K_{gkl})$ and $B_k=\tau b_{gk}$,
respectively.
For a Gaussian distribution written in this canonical form, the mean is given by
\be
\langle x_k \rangle_{\pi_G} = -(A^{-1} B)_k
\label{G1}
\ee
and the covariance matrix by
\be
\langle \qty(x_k-\langle x_k \rangle_{\pi_G})\qty(x_l-\langle x_l \rangle_{\pi_G}) \rangle_{\pi_G} = A^{-1}_{kl}
\label{G2}
\ee
Putting together Eqs.(\ref{moment1},\ref{moment2},\ref{G1},\ref{G2}) the $\tau$-dependent Gaussian parameters are given by
\be
\omega_{gk}(\tau)= \sqrt{\frac{C^{-1}_{kk}(\tau)}{\tau}}
\label{FIT1}
\ee
\be
K_{gkl}(\tau)= \frac{C^{-1}_{kl}(\tau)}{\tau}
\label{FIT2}
\ee
and
\be
b_{gk}(\tau)=- \frac{1}{\tau} \sum_{l=1}^N C^{-1}_{kl}(\tau) \mu_l(\tau)
\label{FIT3}
\ee
where $C$ and $ \mu$ are the exact covariance matrix, 
$C_{kl}(\tau)= \langle \qty(x_k-\langle x_k \rangle_\pi)\qty(x_l-\langle x_l \rangle_\pi) \rangle_\pi$, 
and exact mean, $\mu_l(\tau)= \langle x_l \rangle_\pi$, 
respectively.
Determining the Gaussian parameters thus involves computing the quantities $C_{kl}$ and $\mu_l$,
which together require evaluating $2 N + \frac{N (N-1)}{2}$
integrals over the
$N$-dimensional space, as well as inverting the covariance matrix $C$. 
Here, we assume that $C^{-1}_{kl}$
is positive for all $k$.
This condition is satisfied since the Hessian matrix evaluated at the equilibrium configuration of the oscillators 
has positive eigenvalues.\\

Replacing in the expression of the discretized partition function, Eq.(\ref{Z_path}), the exact potential contribution by its Gaussian approximation,
Eq.(\ref{replaceV}),
we get the following expression for $Z_n$ 
$$
Z_n(\beta)= \qty(\frac{I}{I_G})^n
\qty(\frac{1}{\sqrt{ 2 \pi \tau}})^{nN} \int \prod_{i=1}^n d{\bf x}^i
$$
\be
e^{-\frac{1}{\tau} \sum_{i,j}^n \sum_{k=1}^N x^i_k A_{ij} x^j_k -\tau \sum_{i=1}^n  V_G({\bf x}^i)}.
\label{Zn0}
\ee
The argument of the exponential being a quadratic form in the $nN$ variables, 
the integral can be evaluated analytically.
Let us outline the main steps of this calculation.
Introducing the quantity ${\bar x}_k$ defined as
\be
{\bar x}_k \equiv -\frac{ b_{gk}}{\omega^2_{gk}}
\label{xbar}
\ee
we have
$$
Z_{n}=  \qty(\frac{I}{I_G})^n \qty( \frac{1}{\sqrt{ 2 \pi \tau}})^{nN} 
e^{\frac{\beta}{2} \sum_{k=1}^N \frac{b^2_{gk}}{\omega^2_{gk}}} 
$$
$$
\times \int \prod_{i=1}^N d{\bf x} e^{-\frac{1}{\tau} \sum_{i,j}^n \sum_{k=1}^N x^i_k A_{ij} x^j_k}
$$
$$
\times e^{
	-\tau \qty[ \sum_{i=1}^n \sum_{k=1}^N \frac{1}{2}\omega^2_{gk} x^{i2}_k
 + \sum_{\substack{1 \leq k < l \leq N}} K_{gkl} \qty(x^i_k + {\bar x}_k) \qty( x^i_l + {\bar x}_l) ] }.
$$
Introducing the quantities ${\bar K}_{gk}$ defined as
\be
{\bar K}_{gk} \equiv \sum_{l=1}^N K_{gkl} {\bar x}_l
\label{Kbar}
\ee
$Z_n$ becomes
$$
Z_{n}=  \qty(\frac{I}{I_G})^n \qty( \frac{1}{\sqrt{ 2 \pi \tau}})^{nN} 
e^{ \frac{1}{2} \beta A}
\int d{\bf x}
e^{-P\qty({\bf x}^1,...,{\bf x}^n)}
$$
where the quantity $A$ is given by
\be
A= \sum_{k=1}^N \frac{b^2_{gk}}{\omega^2_{gk}} - 2  \sum_{\substack{1 \leq k < l \leq N}} K_{gkl} {\bar x}_k {\bar x}_l 
\label{bigA}
\ee
and $P$ is a polynomial of order 2 expressed as
$$
P\qty({\bf x}^1,...,{\bf x}^n) = \frac{1}{\tau} \sum_{i,j}^n \sum_{k=1}^N x^i_k A_{ij} x^j_k
$$
$$
+\tau \sum_{i=1}^n \qty( \sum_{k=1}^N \frac{1}{2}\omega^2_{gk} x^{i2}_k
 + \sum_{k=1}^N {\bar K}_{gk} x^i_k + \sum_{\substack{1 \leq k < l \leq N}} K_{gkl} x^i_k x^i_l
  ).
$$
Note that the quadratic form couples only the variables
that share either a common time index $i$
or a common space index $k$.
As a result, the diagonalization of the quadratic form can be carried out independently over the time and space indices.
The "time" rotation consists in diagonalizing the matrix
$A$. The transformed variables are defined as
${\tilde x}^i_k = \sum_{j=1}^n  O^{ij} x^j_k$ where $O^{ij}$
is the transformation matrix. The corresponding eigenvalues are given by
$$
\lambda_i = 1 -\cos{\frac{2\pi}{n}(i-1)} \;\; i=1,n.
$$
For each oscillator $k$ and pair of oscillators $(k,l)$, we have
$$
\sum_{i=1}^n  x^{2i}_k = \sum_{i=1}^n  {\tilde x}^{2i}_k 
$$
$$
\sum_{i=1}^n  x^i_k x^i_l = \sum_{i=1}^n  {\tilde x}^i_k {\tilde x}^i_l,
$$
and
$$
\sum_{i=1}^n  x^i_k = \sum_{i=1}^n \sum_{j=1}^n O^{ji} {\tilde x}^j_k
$$
After some elementary manipulations, we get
$$
Z_{n}=  \qty(\frac{I}{I_G})^n \qty( \frac{1}{\sqrt{ 2 \pi \tau}})^{nN}
e^{ \frac{\beta}{2} A}
\int d{\bf x}
e^{-\sum_{i=1}^n  P_i}
$$
with
$$
P^i= \sum_{k=1}^N \qty(\frac{1}{\tau} \lambda_i +\tau \frac{1}{2}\omega^2_{gk}) x^{i2}_k 
+ \sum_{\substack{1 \leq k < l \leq N}} K_{gkl} x^i_k x^i_l + \sum_{k=1}^N c^i_k x^i_k,
$$
where $c^i_k$ is defined as
$$
c^i_k= \tau {\bar K}_{gk} \sum_{j=1}^n O^{ij}.
$$
For each time coordinate $i$, we now perform the "space" rotation. The matrix $Q^i$ to diagonalize writes
$$
Q^i=\frac{\lambda_i}{\tau} \mathbb{I} + \frac{\tau}{2} M
$$
where $\mathbb{I}$ is the identity matrix and $M$ the symmetric matrix given by
\be
M =
\begin{bmatrix}
\omega^2_{g1} & K_{g12} & \cdots & K_{g1N} \\
K_{g12} & \omega^2_{g2} & \cdots & K_{g2N} \\
\vdots & \vdots & \ddots & \vdots \\
K_{g1N} & K_{g2N} & \cdots & \omega^2_{gN}
\label{matrixM}
\end{bmatrix}
\ee
In the absence of coupling, all eigenvalues of the matrix are positive. 
In physical applications where the coupling terms do not destabilize the system (vibrations around a local minimum), this property remains valid.
Accordingly, we assume throughout that all eigenvalues remain positive.
Let $\Omega^2_{gk}$ 
denote the eigenvalues of the $M$ matrix and
$S_{kl}$ the corresponding transformation matrix.
The transformed variables are expressed as
$$
y^i_k = \sum_{l=1}^N S_{kl} x^i_l.
$$
The eigenvalues of the matrix $Q^i$ will be denoted as $\nu^i_k$; they are given by
$$
\nu^i_k= \frac{1}{\tau} \lambda_i  +  \frac{\tau}{2} \Omega^2_{gk}.
$$
The diagonalization of the quadratic form, $P^i$ leads to
$$
P^i= \sum_{k=1}^N  \nu^i_k y^{i2}_k + u^i_k y^i_{k},
$$
where $u^i_k$ is defined as
$$
u^i_k = \sum_{l=1}^N c^i_{l} \qty(S^T)_{l k}.
$$
After integrating out the variables $y^i_k$, we get 
$$
Z_{n}= \qty(\frac{I}{I_G})^n  e^{ \frac{\beta}{2} A}  e^{ \sum_{i=1}^n   \sum_{k=1}^N  \frac{u^{i2}_k}{4\nu^i_k}}
\prod_{i=1}^n \prod_{k=1}^N \frac{1}{\sqrt{2\tau \nu^i_k}}
$$
In Appendix \ref{appendix_A} the following relation is derived
$$
\sum_{i=1}^n \frac{ \qty(\sum_{j}  O^{ij})^2}{\lambda_i + \tau^2 x} = \frac{n^2}{\beta \tau x}.
$$
Using this equality, we can show that
\be
\sum_{i=1}^n \sum_{k=1}^N  \frac{u^{i2}_k}{4\nu^i_k} = \frac{\beta}{2} \sum_{k=1}^N  \frac{ \qty( \sum_l {\bar K}_{gl} S^T_{lk})^2}{\Omega^2_{gk}}
\label{relation}
\ee
Now, using Eqs.(\ref{bigA}) and (\ref{relation}), the partition function is written as
$$
Z_n= \qty(\frac{I}{I_G})^n  e^{\frac{\beta {\cal A}}{2}}  
\prod_{i=1}^n \prod_{k=1}^N \frac{1}{\sqrt{2\tau \nu^i_k}}
$$
where the quantity ${\cal A}$ is defined as
$$
{\cal A} = \sum_{k=1}^N \frac{b^2_{gk}}{\omega^2_{gk}} - 2 \sum_{\substack{1 \leq k < l \leq N}} K_{gkl} {\bar x}_k {\bar x}_l + \sum_{k=1}^N  \frac{ \qty(\sum_l {\bar K}_{gl} 
S^T_{lk})^2}{\Omega^2_{gk}}.
$$
Quite remarkably, this expression can rewritten in a more compact form. In Appendix \ref{appendix_A} we show that
\be 
{\cal A} = (b_g, M^{-1} b_g)
\ee
where $M$ is the matrix given in Eq.(\ref{matrixM}).
$I_G$ being a Gaussian integral, it can be explicitly evaluated. It is given by
$$
I_G=I_0 e^{ \frac{\tau}{2} \qty(b_g, M^{-1} b_g)}
$$
where 
\be
I_0 = \prod_{k=1}^N \sqrt{ \frac{2\pi}{\tau \Omega^2_{gk}}}.
\label{I0}
\ee
Finally, the discretized partition function given in Eq.(\ref{Zn0}) writes
\be
Z_n =  \qty(\frac{I}{I_0})^n \prod_{i=1}^n \prod_{k=1}^N \frac{1}{\sqrt{2\tau \nu^i_k}}.
\label{Znd}
\ee
Given the somewhat lengthy nature of the calculation, it may be worthwhile to verify numerically the equality of 
the starting expression of $Z_n$, Eq.(\ref{Zn0}) and the final result, Eq.(\ref{Znd}).
This can be easily done by rewriting the original multidimensional integral, Eq.~(\ref{Zn0}), under the form
$\int d{\bf x} e^{- \frac{1}{2} (x, A x) - B x}$ 
where $A$ is an $nN \times nN$ matrix
and $B$ a vector of size $nN$, then by numerically inverting the matrix 
$A$, and, finally, by applying the following standard result for Gaussian integrals
\be
\int d{\bf x} e^{- \frac{1}{2} (x, A x) - B x} =
(\sqrt{2\pi})^{nN} \frac{1}{\sqrt{det A}} e^{\frac{1}{2} (B, A^{-1} B)}.
\ee

The next step is to evaluate the infinite-$n$ limit of the product, 
$\prod_{i=1}^n \frac{1}{\sqrt{2\tau \nu^i_k}}$  in $Z_n$, Eq.(\ref{Znd}). For the moment, the prefactor $\qty(\frac{I}{I_0})^n$
is left unchanged.
Using the following formula (see its derivation in appendix A of \cite{osc1})
$$
\lim_{n \rightarrow \infty} \prod_{i=1}^n \frac{1}{\sqrt{2\lambda_i+\frac{x^2}{n^2}}} = \frac{1}{e^{\frac{x}{2}} - e^{-\frac{x}{2}}}
$$
we get
$$
\lim_{n \rightarrow \infty} \prod_{i=1}^n  \frac{1}{\sqrt{2\tau \nu^k_i}} = \frac{1}{e^{\frac{\beta {\Omega}_{gk} }{2}} 
- e^{-\frac{\beta {\Omega}_{gk}}{2}}}.
$$
The new partition function (denoted as ${\cal Z}$ instead of $Z$) becomes
\be
{\cal Z}_n=  {\cal N}_n \prod_{k=1}^N  \frac{1}{e^{\frac{\beta {\Omega}_{gk} }{2}}
- e^{-\frac{\beta {\Omega}_{gk}}{2}}}
\label{Zn1}
\ee
where we recall that the prefactor ${\cal N}_n$
can be written in two equivalent forms:
\be
{\cal N}_n= \qty(\frac{I}{I_0})^n 
\label{form1}
\ee
or
\be
{\cal N}_n= \qty(\frac{I}{I_G})^n  e^{\frac{{\cal A}}{2}}.
\label{form2}
\ee
Note that both expressions will be used in the following.\\

Taking the limit $n \rightarrow \infty$ for the prefactor ${\cal N}_n$
is not possible, since it does not converge to a finite value, except in the special case of a harmonic potential where
$I=I_0$.
At this point, we introduce the second key idea of this work.
Following what we have done in the one-dimensional case\cite{osc1,osc2}, we propose 
not to take this limit but, instead, {\it for each temperature} to evaluate the partition function at some {\it finite} optimal value of
$n$ (or time step $\tau$). To determine this optimal value, we invoke a
Principle of Minimal Sensitivity (PMS) which imposes
the partition function to be the least sensitive to the variations of the Gaussian parameters
introduced in the Gaussian approximation.
More precisely, for each index $k$ and each pair $(k,l)$, 
we require the following PMS conditions to be satisfied
\be
\frac{\partial {\cal Z}_n}{\partial \omega_{gk}(\tau)}=0,
\label{PMS1}
\ee
\be
\frac{\partial {\cal Z}_n}{\partial K_{gkl}(\tau)}=0.
\label{PMS3}
\ee
Note that the condition $\frac{\partial {\cal Z}_n}{\partial b_{gk}(\tau)} = 0$ is not considered, 
as it is automatically satisfied due to the independence of ${\cal Z}_n$ 
from the vector ${\bf b}_g$; see Eqs. (\ref{Zn1}) and (\ref{form1}).

At this point, it is worth noting that this principle of minimal sensitivity is not new.
It has been introduced in various contexts by several authors (see, {\it e.g.}, \cite{PMS1}, \cite{PMS2}, \cite{PMS3}, \cite{PMS4}).
It is both a general and natural principle: When approximating a quantity by an expression $Q({\bf p})$ (here, ${\cal Z}_n$)
that depends on some {\it unphysical} parameters ${\bf p}$ (here, the Gaussian parameters),
a reasonable criterion for selecting the "best" parameters is to minimize the sensitivity
of the approximation to these parameters, that is, imposing
$\frac{\partial{Q}}{\partial p}=0$.

In the one-dimensional case\cite{osc1,osc2}, there is only one Gaussian parameters $\omega_g$ and the 
PMS condition reduces to
\be
\frac{\partial {\cal Z}_n}{\partial \omega_{g}(\tau)}=0.
\ee 
This condition leads to the following equation for the variable $\tau$
\be
\tau =\frac{2}{ \omega_g(\tau) \coth{\frac{\beta \omega_g(\tau)}{2}}}
\label{PMS1d}
\ee
where the effective frequency, $\omega_g(\tau)$ is given by
\be
\omega_g(\tau)= \frac{1}{\sqrt{\tau \langle \qty(x-\langle x \rangle)^2 \rangle}},
\ee
see Eq.(\ref{FIT1}).
For the quartic oscillator, we have shown\cite{osc1} that the nonlinear PMS equation, Eq.~(\ref{PMS1d}),
admits a unique solution, $\tau(\beta)$ [that is, $n(\beta)$].\\

In the general multi-dimensional case, the PMS conditions can be reformulated as
\be
\tau = \frac{2 \sum_{k=1}^N  \frac{\partial {\Omega}_{gk}(\tau)} {\partial p} \frac{1}{ \Omega_{gk}(\tau)}
}
{
        \sum_{k=1}^N \frac{\partial {\Omega}_{gk}(\tau)} {\partial p} \coth{ \frac{ \beta  \Omega_{gk}(\tau)}{2} }
}
\label{PMSmulti}
\ee
where $p$ denotes one of the Gaussian parameters in the set $\{{\omega}_{gk}(\tau), K_{gkl}(\tau)\}$.
This system of $M = N +\frac{N(N-1)}{2}$ equations [combined with the expressions Eqs.(\ref{FIT1}),(\ref{FIT2}), and (\ref{FIT3}) for the 
Gaussian parameters] is overdetermined and no solution in $\tau$ can be found in general.
To proceed, let us first consider the case of a system of $N$ {\it uncoupled} quantum oscillators. In this case,
the covariance matrix, $\langle (x- \langle x_k \rangle) (x-\langle x_l \rangle) \rangle$  is diagonal, the Gaussian parameters $K_{gkl}$ 
vanish (see, Eq.(\ref{FIT2})), and the partition function
factorizes under the form
\be
{\cal Z}_n= \prod_{k=1}^N z^k_n
\ee
where the $z^k_n$'s are the partition functions of $N$ independent one-dimensional oscillators 
with effective frequency $\omega_{gk}(\tau)$. 
The PMS equations to be solved are now decoupled into $N$ one-dimensional PMS equations, leading 
to $N$ {\it different} optimal time-steps, $\tau_k(\beta)$. 
In presence of couplings the generalization is therefore rather natural: 
A time-step per Gaussian parameter is introduced and the PMS equations become a system of 
$M= N + \frac{N(N-1)}{2}$ equations with $M$ unknowns, a system that may now admit a solution. 
Denoting $\tau_k$ and $\tau_{kl}$ the unknown time-steps associated with $\omega_{gk}$ and
$K_{gkl}$, respectively, the new PMS equations write
\be
\tau_k = \frac{ 2
	\sum_{i=1}^N \frac{\partial {\Omega}_{gi}(\qty[\tau])} {\partial  \omega_{gk}} \frac{1}{ \Omega_{gi}(\qty[\tau])} 
}
{ \sum_{i=1}^N  \frac{\partial {\Omega}_{gi}(\qty[\tau])} {\partial \omega_{gk}}
\coth{ \frac{ \beta  \Omega_{gi}(\qty[\tau])}{2} }}.
\label{solv1}
\ee
and
\be
\tau_{kl} =\frac{ 2
        \sum_{i=1}^N \frac{\partial {\Omega}_{gi}(\qty[\tau])} {\partial K_{gkl}} \frac{1}{ \Omega_{gi}(\qty[\tau])}
} 
{ \sum_{i=1}^N  \frac{\partial {\Omega}_{gi}(\qty[\tau])} {\partial K_{gkl}}
\coth{ \frac{ \beta  \Omega_{gi}(\qty[\tau])}{2} }}.
\label{solv_fin}
\ee
Here, $\qty[\tau]$ denotes the complete set of time-steps, that is, $\qty[\tau] = (\tau_1, \ldots, \tau_N, \tau_{12}, \ldots, \tau_{N-1,N})$.
The existence of a solution to the PMS equations -a system of coupled, nonlinear equations- is a nontrivial mathematical problem. 
Moreover, multiple solutions may exist. In the numerical simulations presented below, which involve simple systems with up to ten coupled oscillators, 
we employed a simple iterative procedure to solve these equations. When initialized with reasonable estimates for the time-steps, 
this method systematically converged. However, such convergence cannot be guaranteed for all types of potentials or larger numbers of oscillators.
A more detailed investigation of this important but challenging issue is left for future work.
Having introduced multiple time-steps, the distribution $\pi$ associated with $e^{-\tau V}$ needs to be generalized. Using Eq.(\ref{V}) for the 
potential, the multiple time-step distribution writes
\be
\pi= \frac{e^{ -\sum_{k=1}^N \tau_k V_k(x_k) - \sum_{\substack{1 \leq k < l \leq N}} \tau_{kl} V_{kl}(x_k,x_l)}}
{ \int d{\bf x} e^{ -\sum_{k=1}^N \tau_k V_k(x_k) - \sum_{\substack{1 \leq k < l \leq N}} \tau_{kl} V_{kl}(x_k,x_l)}}.
\label{pi_multiple}
\ee
The conditions determining the Gaussian parameters, Eqs.(\ref{moment1}) and (\ref{moment2}), must also be generalized.
The new Gaussian parameters are now given by
\be
\omega_{gk}(\qty[\tau])= \sqrt{ \frac{C^{-1}_{kk}(\qty[\tau])}{\tau_k}}
\label{omegagk}
\ee
\be
K_{gkl}(\qty[\tau])= \frac{C^{-1}(\qty[\tau])_{kl}}{\tau_{kl}}
\label{Kgkl}
\ee
and
\be
b_{gk}(\qty[\tau])= - \frac{\qty( C^{-1}(\qty[\tau]) \mu)_k}{\tau_k}
\label{bk}
\ee
where the covariance matrix $C$ and the vector of first-order moments ${\bf \mu}$ are calculated 
using the multiple time-step distribution $\pi$, Eq.(\ref{pi_multiple}).
Note that, in the high-temperature regime, $\beta \rightarrow 0$, $n \sim 1$, and all time-steps become identical with $\tau \sim \beta$, see
Eqs.(\ref{solv1}) and (\ref{solv_fin}). Using Eqs.(\ref{Zn1}) and (\ref{form1}) for the PF, we
get
\be
{\cal Z} \rightarrow \frac{I}{I_0} \prod_{k=1}^N \frac{1}{\beta \Omega_{gk}}
\ee
and using expression (\ref{I0}) for $I_0$
\be
{\cal Z} \rightarrow  \qty(\frac{1}{\sqrt{2\pi\beta}})^N \int d{\bf x} e^{-\beta V({\bf x})},
\ee
which is the exact classical limit.
At finite temperature, the time steps are in general different, and the prefactor
${\cal N}_n$ of ${\cal Z}_n$ becomes ill-defined due to the ambiguity in choosing the unique value of
$n$ (or equivalently, $\tau$).
An additional prescription is therefore required to determine the prefactor.
A first possibility would be to define a single {\it single} effective time step, for instance as a weighted average of the individual time steps
$\tau_k$'s, {\it  i.e.}, ${\bar \tau}=\sum_k w_k \tau_k$.
At small $\beta$ 
where the  $\tau_k$ 
values become similar before converging to the same value at $\beta=0$, 
this approximation is reasonable since the classical limit is recovered. However, at large $\beta$
numerical results indicate that the accuracy of this approximation deteriorates significantly.
After some experimentation, we found that a more robust and accurate strategy is to approximate the prefactor by neglecting 
the coupling terms in the multidimensional integrals $I$ and $I_G$.
To implement this idea, the form (\ref{form2}) of the prefactor is more suitable than (\ref{form1}). 
The approximation we propose therefore reads as follows:
\be
{\cal N}_n = \qty( \frac{I}{I_G})^n  e^{ \frac{\beta}{2} {\cal A}} 
\simeq \prod_{k=1}^N \qty(\frac{I_k(\tau_k)}{I_{Gk}(\qty[\tau])})^{\frac{\beta}{\tau_k}}  e^{ \frac{\beta}{2} {\cal A}(\qty[\tau])}
\label{Nn}
\ee
where
\be
I_k(\tau_k) = \int dx_k e^{-\tau_k V_k(x_k)}
\label{Ik}
\ee
and
\be
I_{Gk}(\qty[\tau]) = \int dx_k e^{-\tau_k \qty( \frac{1}{2} \omega^2_{gk}(\qty[\tau]) x^2_k + b_{gk}(\qty[\tau]) x_k)}.
\label{IGk}
\ee
Note that the $K_{gkl}$
quantities are set to zero only in the multidimensional integral
$I$ and not in ${\cal A}$, which is evaluated exactly. 
Note also that with this choice the classical limit is no longer recovered exactly. However, numerical simulations 
show that the resulting error remains small. In any case, since in this limit the time steps converge to a common value, 
it is not difficult to enforce the exact classical limit by smoothly connecting the approximate multiple–time-step form, Eq.~(\ref{Nn}), 
to the exact single–time-step expression, Eq.~(\ref{form2}).

To summarize, let us present a schematic algorithm for the evaluation of the partition function:\\
\\
i) Start with some initial values for the time-steps, $\qty[\tau]=(\tau_1,...,\tau_N,\tau_{12},...,\tau{N-1 N})$.\\
\\
ii) Evaluate numerically the vector of first-order moments, $\mu_k= \langle x_k \rangle_\pi$, and the covariance matrix 
$C_{kl}= \langle \qty(x_k - \mu_k) \qty(x_l - \mu_l) \rangle_\pi$
with the multiple-step distribution $\pi$, Eq.(\ref{pi_multiple}).\\
\\
iii) Evaluate numerically the inverse of the covariance matrix.\\
\\
iv) Evaluate the new Gaussian parameters, $\omega_{gk}(\qty[\tau])$, $b_{gk}(\qty[\tau])$, for $k=1$ to $N$ and $K_{gkl}(\qty[\tau])$ for 
$k,l=1$ to $N$ with $1 \le k < l \le N$ using Eqs.(\ref{omegagk}),(\ref{Kgkl}), and (\ref{bk}).\\
\\
v) Evaluate the new time-steps using the PMS equations, Eqs.(\ref{solv1},\ref{solv_fin}), where the frequencies $\Omega_{gk}$ are obtained 
by diagonalizing the matrix $M_{kl}=\omega_{gk}(\qty[\tau]) \delta_{kl} + K_{gkl}(\qty[\tau])$, Eq.(\ref{matrixM}). In practice, the derivatives of the 
eigenvalues $\Omega_{gk}$'s with respect to the Gaussian parameters 
can be conveniently performed using a finite-difference scheme.\\
\\
vi) Go to step ii) and iterate up to convergence of the time-steps.\\
\\
vii) Finally, evaluate the partition function using Eqs.(\ref{Zn1}),(\ref{Nn}),(\ref{Ik}), and (\ref{IGk}).

\section{Numerical applications}
\label{applications}
\subsection{Quantum oscillators for $N$=1,2, and 3. Comparison with exact numerical results}
In this section, we consider systems consisting of $N = 1$, 2, and 3 coupled anharmonic oscillators 
and compare our results with "exact" numerical calculations. 
These reference values are obtained by diagonalizing the Hamiltonian matrix in a sufficiently large Gaussian basis set 
to ensure convergence of the energy levels. The partition function is then computed as a sum over a sufficiently large number of components, 
$e^{-\beta E_i}$. We present numerical results for the free energy, $F = -\frac{1}{\beta} \log Z$, the average energy, 
$E= -\frac{\partial \log Z}{\partial \beta}$, and the specific heat, $c_v = \frac{\partial E}{\partial T}$.

The results are illustrated for a Hamiltonian with a potential of quartic polynomial form
$$
H = -\frac{1}{2} \sum_{k=1}^N \frac{\partial^2}{\partial x_k^2} + \sum_{k=1}^N \frac{1}{2} \omega^2_k x^2_k 
$$
\be
+ \lambda \qty(
\sum_{k=1}^N \mu_k x_k^3 + g_k x_k^4 + \sum_{\substack{1 \leq k < l \leq N}} V_{kl})
\ee
where the coupling term is written as
\be
V_{kl}= c^{(1)}_{kl} x_k x_l + c^{(2)}_{kl} x_k^2 x_l + c^{(2)}_{lk} x_l^2 x_k + c^{(3)}_{kl} 
x_k^2 x_l^2
\ee
Here, the parameter $\lambda$
is introduced to study the impact of the anharmonicity and coupling strength on the results. 
In the following applications, the parameters are chosen as follows
$$\omega_k=1+0.2 (k-1)$$
$$\mu_k= -0.25+0.3 (k-1)$$
$$ g_k=1+0.1 (k-1)$$
$$
c^{(1)}_{kl}=0.5 \;\; c^{(2)}_{kl}=0.2 \;\; c^{(2)}_{lk}=0.1 \;\; c^{(3)}_{kl}=0.1 \;\;\; {\rm for} \;\;\; k<l.
$$
Note that we have chosen $c^{(2)}_{kl} \ne c^{(2)}_{lk}$ to consider the general case of a non-symmetric coupling matrix $V_{kl}$.
Calculating the average energy and specific heat requires evaluating the first and second derivatives of the partition function with respect 
to the inverse temperature $\beta$. While it is, in principle, possible to derive explicit analytical expressions for these derivatives, the 
resulting formulas tend to be quite involved. This complexity arises primarily from the need to differentiate the eigenvalues $\Omega_k$. 
In practice, a simpler and computationally efficient approach is to approximate the derivatives using finite-difference methods. We use the following
formulas
$$
E= - \frac{ \log{Z(\beta + \epsilon)}- \log{Z(\beta - \epsilon)}}{ 2 \epsilon}
$$
and
$$
c_v= \beta^2 \frac{ \log{Z(\beta + \epsilon)}- 2 \log{Z(\beta)}+ \log{Z(\beta - \epsilon)}}{\epsilon^2}
$$
where $\epsilon$ denotes a small variation in $\beta$. In practice, we have observed that the computed quantities stabilize rapidly as $\epsilon$ is 
decreased, confirming that the finite-difference approach proposed here is appropriate.
From a numerical point of view, the most computationally demanding step is the evaluation, at each iteration, of 
the $N$-dimensional integrals involved in the formulas giving the components of the first-order moment vector and of the covariance matrix.
For the small number of oscillators considered in this section, this is not an issue, 
as a simple Riemann integration scheme on a sufficiently fine uniform grid is more than suffficient. 
The treatment of much larger values of $N$ is more demanding and will be discussed in the next section.
Tables \ref{T1}, \ref{T2}, and \ref{T3} present a comparison between the results obtained using our analytical partition function 
and exact numerical results 
for systems of one, two, and three oscillators, respectively. Each table reports the free energy, average energy, and specific heat -each expressed 
per oscillator. Results are provided for three selected inverse temperatures, $\beta=0.5, 1.$ and 2., 
and for three values of the anharmonicity/coupling parameter $\lambda=0.1,1,$ and 2.
At the smallest inverse temperature considered ({\it i.e.}, highest temperature), $\beta=0.4$,
and for the weakest anharmonicity and coupling, $\lambda=0.1$,
the relative errors for all quantities are particularly small -typically less than 1\%, with a maximum of 1.5\% for $N=3$.
This high level of accuracy is not surprising since our approximate partition function becomes exact in the limit $\lambda \rightarrow 0$ 
and is also nearly exact as $\beta \rightarrow 0$ (Recall that, for $N \geq 2$, there is a small residual error due to the approximation 
of the ratio of integrals, Eq.(\ref{Nn})).
As the temperature decreases, the relative errors increase slightly but remain small -typically within a few percent. 
Note that the errors in the various quantities at $\beta=1$ and $\beta=2$
are similar, and this trend persists for larger values of
$\beta$ (not shown), indicating no degradation of accuracy at lower temperatures. 
As expected, the errors increase with stronger anharmonicity and coupling, but they remain modest, again within a few percent. 
A slight increase in error is also observed as the number of oscillators increases; 
however, given the small system sizes considered here, it is difficult to draw a firm conclusion regarding a general trend. 
This point will be further explored in the next section, where larger systems are investigated.
Overall, we conclude that for the small systems considered in this section, our approximate analytical partition function yields quite accurate results over a broad range of parameters.

\begin{table*}[htb]
\centering
\caption{$N=1$. Comparison of the analytical and exact results for the thermodynamic quantities $F/N,E/N$, 
	and $c_v/N$ at three values of the inverse temperature $\beta$ and coupling parameter $\lambda$. 
	Relative errors $\epsilon$ (in $\%$) on each quantity $Q$ is reported,
	$\epsilon={(Q_{approx}-Q_{ex})}/{Q_{ex}} \times 100$.}
\label{T1}
\begin{tabular}{cccccccccc}
\hline
\multicolumn{4}{c}{$\beta = 0.5$} & \multicolumn{3}{c}{$\beta = 1$} & \multicolumn{3}{c}{$\beta = 2$} \\
	& Analytical formula & Exact & $\epsilon(\%)$ & Analytical formula & Exact &  $\epsilon(\%)$ & Analytical formula & Exact &  $\epsilon(\%)$ \\
$\lambda=0.1$ &  &  &  &  &  &  &  &  &  \\
 $F/N$ &  -0.88533  &  -0.88174 &   0.4 &   0.22177 &   0.22530  &   -1.5  &  0.51023  &   0.51336     &  -0.6 \\
 $E/N$ &    1.82260 &    1.82620&  -0.1&    1.02293  &    1.02605  &  -0.3   &  0.66944  &   0.67188 &  -0.3    \\
 $c_v/N$ &   0.81039      &   0.81051  &   -0.01  &   0.77154  &   0.77275   &  -0.1  &  0.58292 &   0.58383    &  -0.1   \\
	\\
$\lambda=1$ &  &  &  &  &  &  &  &  &  \\
	$F/N$ &   -0.07155  &   -0.05308  &    34.7$^{a}$  &   0.63108 &   0.65114    &   -3.0 &  0.76903 &   0.78853    &   -2.4  \\
 $E/N$ &    1.76696    &    1.78923  &   -1.2  &    1.07432 &    1.09472&   -1.8 &  0.82153  &   0.83995    &   -2.1 \\
	$c_v/N$ &   0.72839 &   0.72810  &  0.04 &   0.62244 &   0.62754&  -0.8  &  0.31544  &   0.31443    &   0.3   \\
\\
$\lambda=2$ &  &  &  &  &  &  &  &  &  \\
 $F/N$ &   0.24716  &   0.27430 &   -9.8 &   0.81631  &   0.84537   &   -3.4   &  0.90999   &   0.93837   &   -3.0   \\
 $E/N$ &    1.79690  &    1.82933  &   -1.7    &    1.14493  &    1.17396 &   -2.4  &  0.93770   &   0.96524  &   -2.8 \\
 $c_v/N$ &   0.70239 &   0.70277 &   -0.05 &   0.55615  &   0.56267   &   -1.1 &  0.20823&   0.20523  &    1.4  \\
\hline
\end{tabular}
\footnotetext[\value{footnote}]{$^{a}$The relative error is accidentally large because $F/N$ is small. 
This is not the case of the absolute error.}
\end{table*}

\begin{table*}[htb]
\centering
	\caption{$N=2$.
	Comparison of the analytical and exact results for the thermodynamic quantities $F/N,E/N$,
        and $c_v/N$ at three values of the inverse temperature $\beta$ and coupling parameter $\lambda$.
        Relative errors $\epsilon$ (in $\%$) on each quantity $Q$ is reported,
        $\epsilon={(Q_{approx}-Q_{ex})}/{Q_{ex}} \times 100$.}
\label{T2}
\begin{tabular}{cccccccccc}
\hline
\multicolumn{4}{c}{$\beta = 0.5$} & \multicolumn{3}{c}{$\beta = 1$} & \multicolumn{3}{c}{$\beta = 2$} \\
        & Analytical formula & Exact & $\epsilon(\%)$ & Analytical formula & Exact &  $\epsilon(\%)$ & Analytical formula & Exact &  $\epsilon(\%)$ \\
$\lambda=0.1$ &  &  &  &  &  &  &  &  &  \\
 $F/N$ &  -0.76924  &  -0.76164 &   0.9 &   0.29453 &   0.29941  &   -1.6 &  0.56197 &   0.56555  &  -0.6    \\
 $E/N$ &    1.85611&    1.85777&   -0.08  &    1.05101  &    1.05325   &  -0.2  &  0.70362 &   0.70596  &  -0.3  \\
$c_v/N$ &   0.81893 &   0.81813&    0.09 &   0.77034 &   0.77025 &   0.01 &  0.55403  & 0.55308 &   0.1  \\
\\
$\lambda=1$ &  &  &  &  &  &  &  &  &  \\
 $F/N$ &   -0.0052  &    0.00209 &   -125.2$^{a}$&   0.67649&   0.70023&   -3.3&  0.80700  &   0.82876&   -2.6   \\
 $E/N$ &    1.79059 &    1.81194  &   -1.1 &    1.10069 &    1.12119  &   -1.8 &  0.85541 &   0.87524 &   -2.2 \\
$c_v/N$ &   0.72907 &  0.72816 &   0.1&   0.61345&   0.61672&  -0.5&  0.29649  &   0.292335451  &    1.4 \\
\\
$\lambda=2$ &  &  &  &  &  &  &  &  &  \\
 $F/N$ &   0.30146   &   0.33755 &   -10.6&   0.85463 &   0.88860 &   -3.8 &  0.94452&   0.97632  &   -3.2  \\
 $E/N$ &    1.81591   &    1.84862  &   -1.7 &    1.17084 &    1.20121&   -2.5 &  0.97127&    1.00130  &   -2.9 \\
 $c_v/N$ &   0.69994&   0.69953 &    0.05  &   0.54242 &   0.54748  &  -0.9 &  0.19784 &   0.19119&    3.4    \\
\hline
\end{tabular}
\footnotetext[\value{footnote}]{$^{a)}$The relative error is accidently large because $F/N$ is small.
This is not the case of the absolute error.}
\end{table*}

\begin{table*}[htb]
\centering
\caption{$N=3$. Comparison of the analytical and exact results for the thermodynamic quantities $F/N,E/N$,                  
and $c_v/N$ at three values of the inverse temperature $\beta$ and coupling parameter $\lambda$.
        Relative errors $\epsilon$ (in $\%$) on each quantity $Q$ is reported,
        $\epsilon={(Q_{approx}-Q_{ex})}/{Q_{ex}} \times 100$.}
\label{T3}
\begin{tabular}{cccccccccc}
\hline
\multicolumn{4}{c}{$\beta = 0.5$} & \multicolumn{3}{c}{$\beta = 1$} & \multicolumn{3}{c}{$\beta = 2$} \\
        & Analytical formula & Exact & $\epsilon(\%)$ & Analytical formula & Exact &  $\epsilon(\%)$ & Analytical formula & Exact &  $\epsilon(\%)$ \\
$\lambda=0.1$ &  &  &  &  &  &  &  &  &  \\
	$F/N$ &  -0.65609  &  -0.64610   &    1.5 &   0.36535 &  0.37092&   -1.5 &  0.61321 &   0.61701 &  -0.6   \\
 $E/N$ &    1.88765&    1.88758&    0.03&    1.07902&    1.08066  &  -0.1  &  0.73936 &   0.74169&  -0.3    \\
 $c_v/N$ &   0.82630 &   0.82386&   0.2 &   0.76617&   0.76501&   0.1&  0.52290   &   0.52076 &   0.4  \\
\\
$\lambda=1$ &  &  &  &  &  &  &  &  &  \\
 $F/N$ &    0.05707 &    0.08642  &   -33.9$^{a}$  &   0.71980&   0.74432&   -3.2 &  0.84354&   0.86540   &   -2.5  \\
 $E/N$ &    1.81391&    1.83301 &   -1.0 &    1.12669  &    1.14606  &   -1.6 &  0.88818 &   0.90779  &   -2.1 \\
$c_v/N$ &   0.72951 &   0.72743  & 0.2 &   0.60509 &   0.60693  &  -0.3  &  0.27975 &   0.27380&    2.1   \\
\\
$\lambda=2$ &  &  &  &  &  &  &  &  &  \\
 $F/N$ &   0.34997   &   0.38834     &   -9.8 &   0.88935 &   0.92355 &  -3.7      &  0.97555  &    1.00700   &   -3.1    \\
$E/N$ &    1.83413  &    1.86428  &   -1.6 &    1.19456  &  1.22352 &   -2.3  &   1.00085 &    1.03027  &   -2.8    \\
 $c_v/N$ &   0.69776  &   0.69575   &   0.2 &   0.53199&   0.53592&  -0.7  &  0.18852 &   0.18026 &    4.5 \\
\hline
\end{tabular}
\footnotetext[\value{footnote}]{$^{a)}$The relative error is accidently large because $F/N$ is small.
This is not the case of the absolute error.}
\end{table*}

\subsection{Quantum oscillators for $N > 3$. Comparison with PIMC}
In this section, we study the applicability and precision of our analytical formula for systems with larger
numbers of oscillators.  To this end, we consider a system of
$N$ coupled quartic oscillators described by the Hamiltonian
\be
H = -\frac{1}{2} \sum_{k=1}^N  \frac{\partial^2}{\partial x^2_k} + \sum_{k=1}^N \frac{1}{2} \omega^2 x^2_k + g x^4_k
+ K \sum_{k=1}^N x_k x_{k+1}
\label{HQ}
\ee
with $x_{N+1}=x_N$ (periodic conditions). Physically, this Hamiltonian can describe, for instance, a chain of coupled oscillators in solid-state physics\cite{ashcroft1976solid},
or it may arise from the discretization of a quantum field theory, such as the
$\phi^4$-theory\cite{ramond1990field}.
In the applications to follow, the Hamiltonian parameters are fixed as $\omega=1$, $g=1$, and $K=0.8$.
As the number of oscillators becomes large, it becomes increasingly difficult 
to obtain "exact" thermodynamic quantities from numerical diagonalization.
To assess the accuracy of our analytical approximation for large systems, we thus compare our results to data obtained 
with Path Integral Monte Carlo (PIMC).

Thanks to the symmetry of the Hamiltonian (translational symmetry and oscillators with identical frequency $\omega$ and coupling constant $g$)
the first-order
moments vanish and the covariance matrix $C_{kl}$ depends only on the distance, $d(k,l)$, between $k$ and $l$ on the circle
of length $N$, that is
\be
C_{kl} = {\cal C}\qty[d\qty(k,l)]
\ee
where $d(k,l)$ is given by
\be
d(k,l)={\rm Min}(|k-l|,N-|k-l|),
\ee
and ${\cal C}$ is some function to determine. Mathematically, it means that the covariance matrix is a symmetric circulant matrix\cite{davis1994circulant}. Here, we shall write it under the following form
\be
C = \begin{bmatrix}
	C_{[0]}     & C_{[1]} & \cdots & C_{[2]}    & C_{[1]}    \\
	C_{[1]}     & C_{[0]} & C_{[1]}    &        & C_{[2]}    \\
\vdots  & C_{[1]} & C_{[0]}    & \ddots & \vdots \\
C_{[2]}     &     & \ddots & \ddots & C_{[1]}    \\
	C_{[1]}     & C_{[2]} & \cdots & C_{[1]} & C_{[0]}    \\
\end{bmatrix},
\ee
with the notation 
\be
C_{[m]} \equiv {\cal C}\qty[d\qty(k,l)=m],
\label{Cm}
\ee
where the distance $m$ varies from $m=0$ to $\lfloor \frac{N}{2} \rfloor$, the symbol $\lfloor ... \rfloor$ denoting the euclidean division.
The notation introduced in Eq.(\ref{Cm}) will be used for all symmetric circulant matrices to follow.
Due to i) the structure of the PMS equations, Eqs.(\ref{solv1},\ref{solv_fin}), ii) the relations giving the Gaussian 
parameters, 
Eqs.(\ref{omegagk},\ref{Kgkl},\ref{bk}) and iii) the fact that the inverse of a symmetric circulant matrix 
is a symmetric circulant matrix, we have the following properties:

i) The $\omega_{gk}$'s and $\tau_k$'s are independent of $k$

ii) The matrices $K_{gkl}$, $M_{kl}$, and $\tau_{kl}$ are symmetric circulant matrices.

Finally, let us note that the spectrum of a circulant matrix is analytically known. For the matrix $M$ 
the eigenvalues $\Omega_{gi}$ are as follows.

For $N$ even
\be
\Omega_{gi}= \sqrt{\omega^2_g + K_{[\frac{N}{2}]} (-1)^i + 2 \sum_{j=1}^{\frac{N}{2}-1} K_{[j]} \cos{\qty(\frac{2\pi}{N} ji)}}  \,\; i=1,N.
\ee
For $N$ odd
\be
\Omega_{gi}= \sqrt{\omega^2_g + 2 \sum_{j=1}^{\frac{N-1}{2}} K_{[j]} \cos{\qty(\frac{2\pi}{N} ji)}} \;\; i=1,N.
\ee
where 
\be
\omega_{g}= \sqrt{\frac{C^{-1}_{[0]}}{\tau_{[0]}}}
\ee
and
\be
K_{[m]}= \frac{C^{-1}_{[m]}}{\tau_{[m]}} \;\;\; m=1,\lfloor \frac{N}{2} \rfloor
\label{FIT22}
\ee
In these relations, $\tau_{[0]}=\tau_k$ and $\tau_{[m]}=\tau_{kl}$.\\

Finally, the PMS equations to solve write in the form
\be
\tau_{[m]} =\frac{ 2
	\sum_{i=1}^N \cos{\qty(\frac{2\pi}{N}mi)} \frac{1}{ \Omega^2_{gi}}
}
{\sum_{i=1}^N  \cos{\qty(\frac{2\pi}{N}mi)} \frac{1}{\Omega_{gi}}   
\coth{\frac{\beta \Omega_{gi}}{2} }}   \;\; m=0,\lfloor \frac{N}{2} \rfloor
\label{solv_fin2}
\ee

As already pointed out in the previous section, 
the most computationally demanding step is the evaluation, at each iteration, of the covariance matrix. 
This requires the calculation of
$$
M = N + \frac{N(N+1)}{2}
$$
$N$-dimensional integrals corresponding to the components of the first-order moment vector [Eq.~(\ref{moment1})] 
and of the covariance matrix, whose elements are given here by
\begin{equation}
C_{kl}=
\frac{
\int d{\bf x}\; x_k x_l
e^{-\tau_{[0]} \sum_{i=1}^N V_i(x_i)-\tau_{[1]} K \sum_{i=1}^N x_i x_{i+1}}
}
{
\int d{\bf x} \;
e^{-\tau_{[0]} \sum_{i=1}^N V_i(x_i)-\tau_{[1]} K \sum_{i=1}^N x_i x_{i+1}}
}.
\end{equation}
For the $N=1$, 2, and 3 cases discussed in the previous section, these integrals can be evaluated 
essentially instantaneously. However, the computational cost grows rapidly with increasing $N$. 
For larger systems (say, $N \gtrsim 10$), it is likely that Monte Carlo methods 
combined with some correlated sampling technique for parameter updates, will become necessary.

In the present work, however, systems containing up to ten coupled oscillators can still be treated 
using a deterministic approach. To this end, the required integrals are evaluated through 
a systematic expansion of the interaction term in the potential, performed until convergence is achieved. 
The expansion reads
\be
e^{-\tau_{[1]} K \sum_{i=1}^N x_i x_{i+1}} = \sum_{p=0}^\infty \frac{\qty(-\tau_{[1]} K)^p}{p!} \qty(\sum_{i=1}^N x_i x_{i+1})^p.
\ee
Using the formula
\be
\qty(\sum_{i=1}^N x_i x_{i+1})^p =\sum_{k_1 +...+k_N=p \;\; k_i \geq 0} \frac{ p!} {k_1! ... k_N!} \prod_{i=1}^N x^{ k_i+ k_{i+1}}_i
\ee
with $x_{N+1}=x_1$ and $k_{N+1}=k_1$, we get
\begin{widetext}
	\begin{equation}
C_{kl} =\frac{ \sum_{p=0}^\infty \qty(-K\tau_{[1]})^p \sum_{k_1 +...+k_N=p \;\; k_i \geq 0}  \frac{ 1} {k_1! ... k_N!}
	\prod_{i=1}^N J_{k_i+k_{i+1}+ \delta_{ik} + \delta_{il}}(\tau_{[1]},V_i)
	}
	{ \sum_{p=0}^\infty \qty(-K\tau_{[1]})^p \sum_{k_1 +...+k_N=p \;\; k_i \geq 0}  \frac{ 1} {k_1! ... k_N!}
        \prod_{i=1}^N J_{k_i+k_{i+1}}(\tau_{[1]},V_i)
        }
	\end{equation}
\end{widetext}
where we have introduced the following one-dimensional integrals
\be
J_n(\tau_{[1]},V_i)= \int dx e^{-\tau_{[1]} V_i(x)} x^n.
\ee
The number of partitions of the integer
$p$ grows exponentially, making this approach feasible only for a relatively small number of
oscillators. In practice, the convergence of the sums as a function of
$p$ can be achieved with sufficient accuracy for values up to approximately $N \sim 10$. To accelerate the convergence of the partial sums 
a Shank's transform\cite{Shanks_1955} is used.
For larger systems, deterministic computations become impractical, and stochastic methods must be considered. 
However, care must be taken due to the potentially severe impact of statistical noise on the iterative scheme.
To circumvent this issue, a possibility is to introduce a correlated sampling strategy, that is, to use a fixed set of configurations 
and to introduce some reweighting in the stochastic averages. 
We have implemented such a Monte Carlo approach and verified that it may work for larger values of $N$.
However, we have observed that the stability of the iterative algorithm, in its simplest form, is no longer systematically maintained.
This problem calls for further refinement and the development of more robust algorithms. A detailed investigation of these improvements 
is beyond the scope of the present work and is deferred to future research.

\begin{figure}[H]
\centering
\includegraphics[scale=0.45]{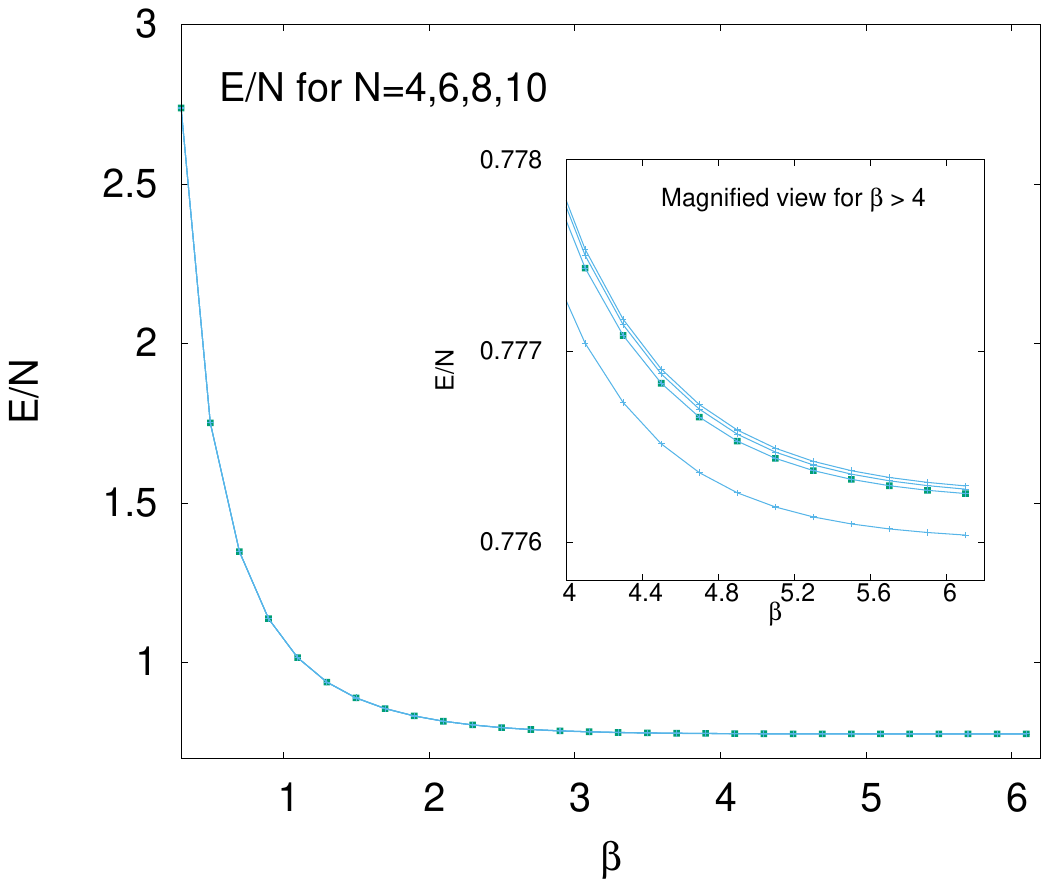}
        \caption{
Average energy per oscillator $E/N$ as a function of $\beta$ for $N=4,6,8$, and $10$. 
At the scale of the main figure, the 
curves for different $N$ almost coincide. The inset shows a magnified view of the curves for $\beta > 4$. 
	From bottom to top, the curves correspond to $N=4$, $N=6$, $N=8$, and $N=10$, respectively.
        }
        \label{fig1}
\end{figure}

\begin{figure}[H]
\centering
\includegraphics[scale=0.45]{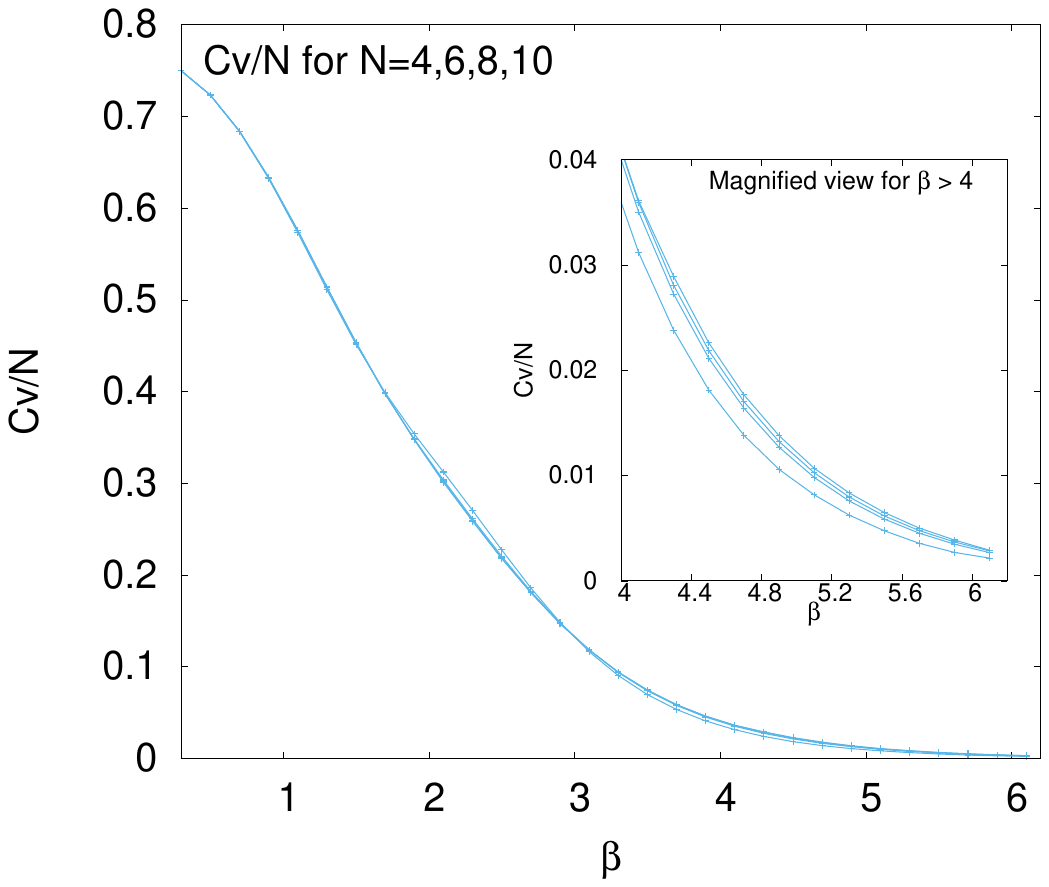}
        \caption{
		Specific heat per oscillator $c_v/N$ as a function of $\beta$ for $N=4,6,8$, and $10$. At the scale of the main figure, the
            curves for different $N$ almost coincide. The inset shows a magnified view of the curves for $\beta > 4$. 
	From bottom to top, the curves correspond to $N=4$, $N=6$, $N=8$, and $N=10$, respectively.
        }
        \label{fig2}
\end{figure}

\begin{figure}[H]
\centering
\includegraphics[scale=0.4]{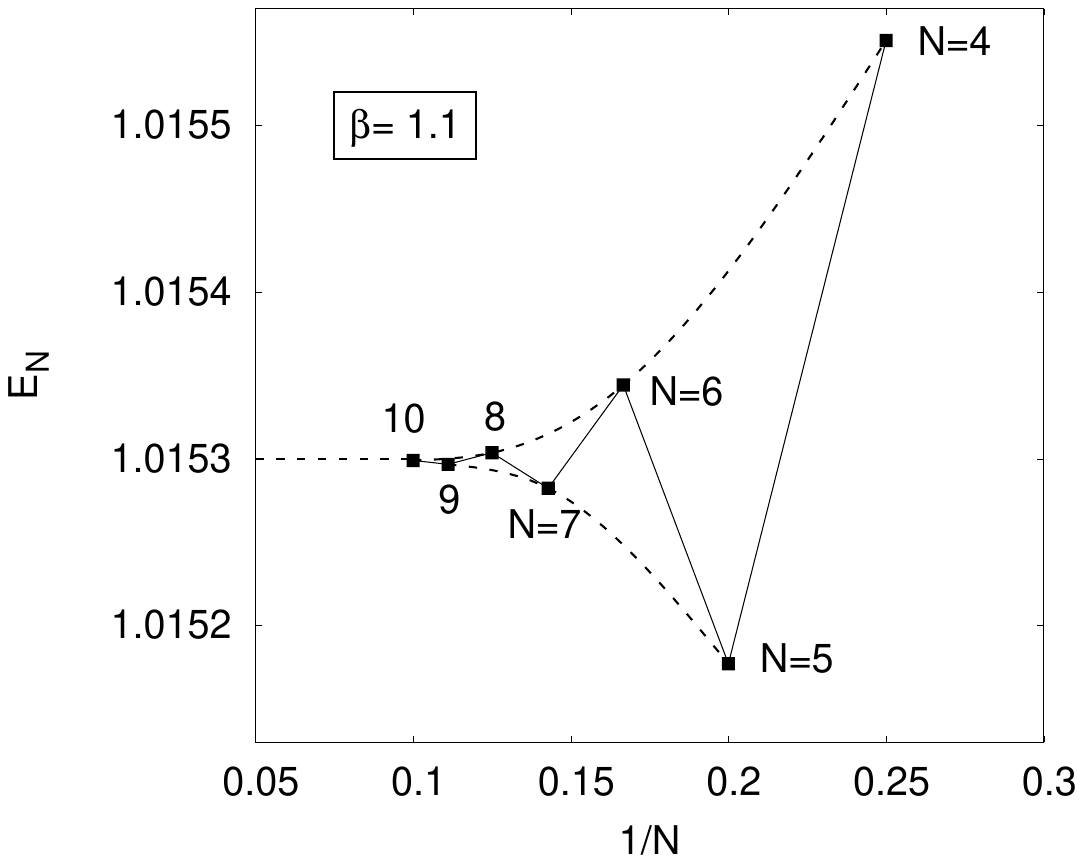}
        \caption{
		$N$-dependence of the average energy per oscillator $E/N$ as a function of $N$ for $\beta=1.1$. For both $N$ even and odd, 
	 a dashed line is drawn for guiding the eye.
        }
        \label{fig3}
\end{figure}

In Figs \ref{fig1} and \ref{fig2} we show the average energy and specific heat per oscillator as a function of $\beta$ obtained 
with our analytical formula. Four curves corresponding to $N=4,6,8$ and 10 are depicted. As seen from the figures, the various curves almost 
coincide at the scale of the main figure, meaning that the finite-size effects are small. To have a closer look at these effects, an inset 
giving a magnified view of the curves at low temperatures, $\beta > 4$, is given for both figures.
To solve the PMS equations, Eq.(\ref{solv_fin2}), we use a simple iterative process in which the different time-steps are evolved as follows
\be
\tau^{n+1}_{[m]} = f_m\qty(\tau^{n}_{[1]}, \tau^{n}_{[2]}, ...)
\ee
where the superscript $n$ is the iteration index and $f_m$ is given by the R.H.S. of Eq.(\ref{solv_fin2}).

For all $\beta$ and number of oscillators considered here ($N \le 10$) we have found that the iterative process 
is remarkably stable and 
converges rapidly in a few iterations to a unique set of time-steps, provided reasonable starting values, $\tau^{0}_{[m]}$, are chosen.
In addition, both thermodynamic quantities are found to display a regular 
behavior when varying $\beta$ and $N$, a satisfactory result for our approximate formula. 
In figure \ref{fig3} we provide a quantitative assessment of $N$-dependence of the average energy as a function of the number $N$ of 
oscillators. The results are shown for a particular value of the inverse temperature, $\beta=1.1$, but
similar behaviors are obtained for all temperatures and for the specific heat too. 
As seen on the figure, a parity effect is observed as a function of $N$.
However, both energy curves corresponding to $N$ even and odd converge rapidly and smoothly to a common value for $N=\infty$. 

To assess the accuracy of our approximate formula, we compare our results with "exact" numerical data obtained using Path Integral Monte 
Carlo (PIMC) simulations. Path integral methods are, in principle, numerically exact -provided the two primary sources 
of error are properly controlled.
The first source of error is the systematic bias associated with using a finite number of time slices, or Trotter number, $n$.
In practice, simulations are carried out for several values of $n$,
and results are extrapolated to the limit $n \rightarrow \infty$.
At high temperatures, the PIMC time step, $\tau=\frac{\beta}{n}$,
can be made sufficiently small with moderate values of
$n$, making the extrapolation relatively straightforward. However, at low temperatures, larger values of
$n$ are required, leading to more computationally intensive simulations.
In addition to the systematic bias, one must account for statistical errors inherent to any stochastic (Monte Carlo) method. 
For each value of $n$,
the statistical uncertainty must be reduced enough to enable a reliable extrapolation of thermodynamic properties to
$n \rightarrow \infty$.
Various PIMC implementations exist in the literature, each with different strengths in terms of efficiency and accuracy. 
However, our objective here is not to employ the most advanced algorithm, but rather to evaluate the accuracy of our formula. 
To this end, we have implemented a standard PIMC approach based on Fourier modes and low-variance estimators for both the average energy 
and specific heat. The relevant expressions are detailed in Appendix \ref{appendix_B}.
To ensure a meaningful comparison with our formula, the PIMC simulations were run long enough so that statistical errors 
are negligible on the scale of our figures (except for the specific heat data shown in the magnified view of Figure \ref{fig5} ). As is well known, the finite-$n$ bias becomes more pronounced as temperature decreases\cite{Uhl_2016}.
Here, this effect is particularly significant in the specific heat at low temperatures (see Figure \ref{fig5}). 
More sophisticated path-integral techniques -such as Path Integral Molecular Dynamics (PIMD) with 
optimized thermostats\cite{Ceriotti_2010, Uhl_2016}. 
could significantly reduce this bias. However, as already noted, such refinements are not critical here for our purpose.

\begin{figure}[H]
\centering
\includegraphics[scale=0.4]{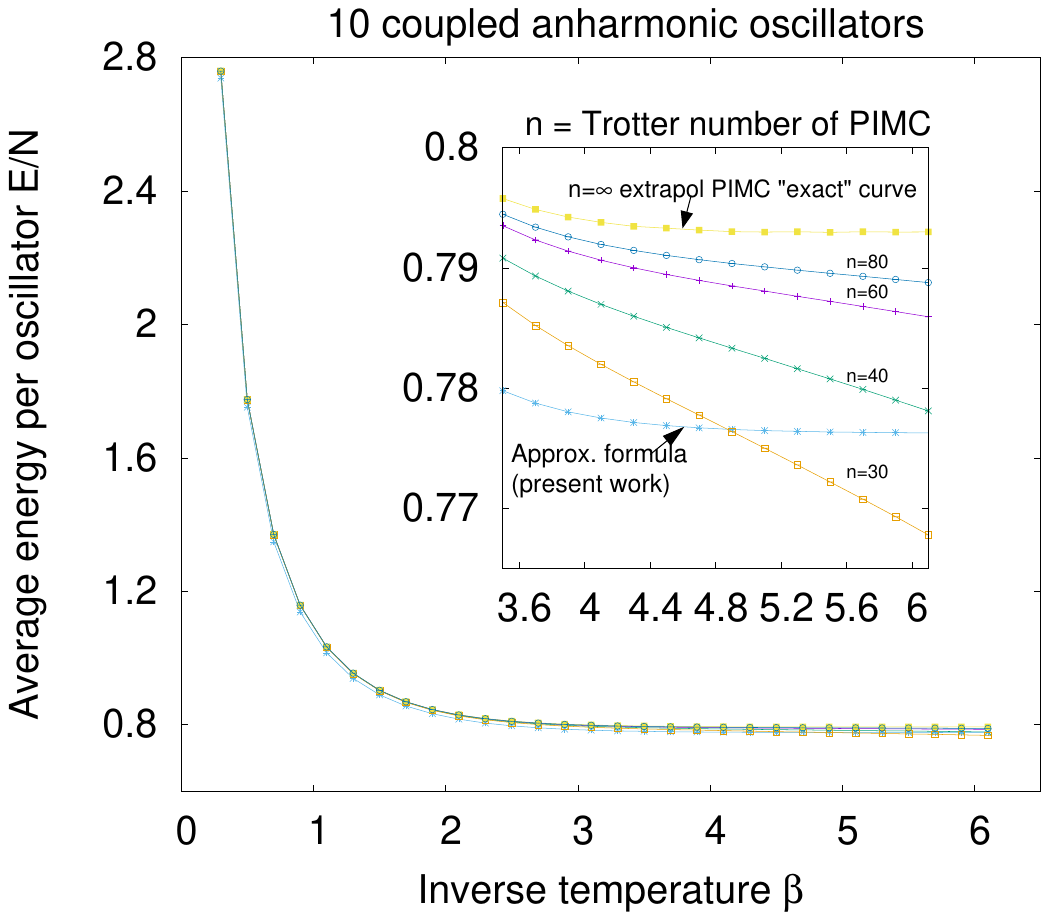}
        \caption{
$N=10$.	Average energy $E/N$ as a function of $\beta$ obtained with PIMC and compared with that obtained with our analytical formula for the 
	partition function. Statistical errors on PIMC data are smaller than point symbols. The PIMC energy curves for $n=30,40,60$ and 80 are displayed together with the extrapolated curve at 
	$n=\infty$. The inset is a magnified view of the low-temperature regime. 
	}
	\label{fig4}
\end{figure}

\begin{figure}[H]
\centering
\includegraphics[scale=0.4]{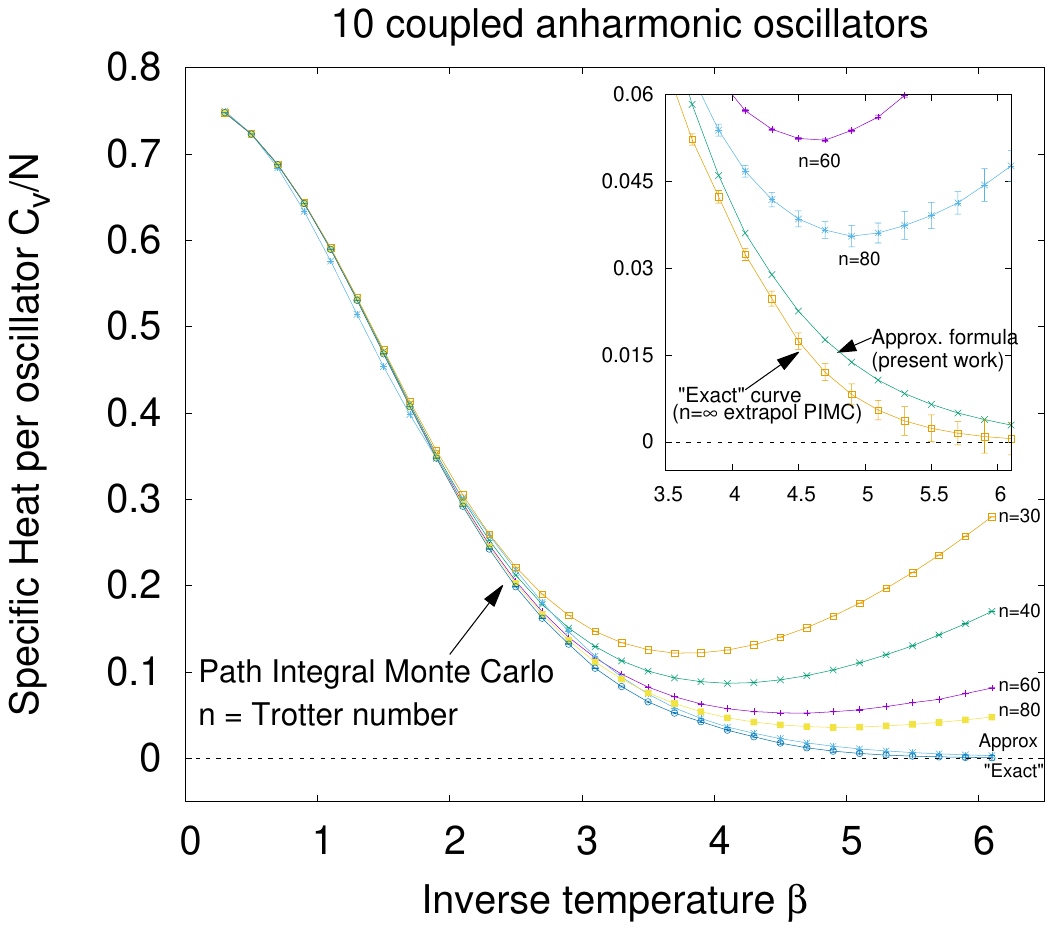}
        \caption{
$N=10$. Specific heat $c_v/N$ as a function of $\beta$ obtained with PIMC and compared with that obtained with our analytical formula for the
        partition function. On the main graph, statistical errors on PIMC data are smaller than point symbols. The PIMC energy curves for $n=30,40,60$ and 80 are displayed together with the extrapolated curve at
        $n=\infty$. The inset shows a magnified view of the low-temperature regime where the approximate and exact PIMC curves can be better
	distinguished.
        }
	\label{fig5}
\end{figure}

Figures \ref{fig4} and \ref{fig5} display the average energy and specific heat per oscillator, respectively, as functions of $\beta$ for $N = 10$. 
The PIMC results for $n = 30$, $40$, $60$, and $80$ are shown together with the curves obtained using our approximate formula for comparison. 

For the average energy, the finite-$n$ biases are relatively small for all temperatures at the scale of the figure, 
and our approximate curve closely follows the PIMC results. A magnified view of the curves is provided in the inset, 
where we focus on the low-temperature regime ($\beta > 3.5$), where the approximation error of our formula is expected to be maximal. 
The PIMC results for various $n$ values have been extrapolated to $n \to \infty$ using a quadratic fit. Notably, individual PIMC curves 
at finite $n$ exhibit a completely unphysical behavior as a function of the temperature: the expected saturation behavior at small temperature 
is absent and,
instead, a spurious linear decreasing behavior is observed. Only, the extrapolated PIMC curve displays 
the correct asymptotic behavior. Remarkably, our approximate expression for the average energy correctly captures this saturation.
Furthermore, the curve remains remarkably 
parallel to the extrapolated exact PIMC curve over the entire temperature range. This is a very satisfactory result since physically 
we are in most cases interested in the global behavior of the energy and not its absolute value.

For the average energy, the finite-$n$ biases are relatively small at all temperatures on the scale of the figure, 
and our approximate curve closely follows the PIMC results. A magnified view is provided in the inset, 
focusing on the low-temperature regime ($\beta > 3.5$), where the approximation error of our formula 
is expected to be maximal. The PIMC results obtained for various
$n$ values have been extrapolated to the
$n \to \infty$ limit using a quadratic fit.
As seen, the individual PIMC curves at finite
$n$ exhibit a completely unphysical temperature dependence: the expected saturation behavior at low temperature is absent 
and is replaced by a spurious linear decrease. Only the extrapolated PIMC curve recovers the correct asymptotic behavior. 
Remarkably, our approximate expression for the average energy correctly captures this saturation. Furthermore, the resulting curve 
remains nearly parallel to the extrapolated PIMC curve over the entire temperature range. This is a very satisfactory result, 
since in most cases the physically relevant information lies in the overall behavior of the energy rather than in its absolute value.

For the specific heat (Fig. \ref{fig5}), a stronger dependence on the Trotter number $n$ is observed at low temperatures. 
As in the case of the average energy, 
extrapolation of the PIMC data yields a well-defined "exact" numerical curve that correctly converges to zero in the low-temperature limit. 
In sharp contrast, at finite $n$  
the PIMC curves exhibit an unphysical minimum and increase without bound as the temperature decreases, which is clearly incorrect.
As expected from the nearly-independent error in the approximate energy as a function of the temperature already mentioned above
(see, Fig. \ref{fig4}) 
and the fact that the specific heat is defined as the derivative of the energy, our formula yields 
a remarkably accurate approximatiom for $c_v$. On the scale of the main figure, the exact and approximate curves are almost indistinguishable. 
The inset provides a closer look at this result.
We emphasize that the level of accuracy achieved here using our analytical partition function is particularly remarkable.

\section{Conclusions}
\label{conclusion}
In this work, we have derived a general approximate expression for the quantum partition function of an arbitrary number 
of coupled quantum oscillators. In practice, evaluating this expression requires solving a system of nonlinear coupled equations, 
the size of which scales with the number of parameters entering the potential energy function. For systems involving up to
$N=10$ coupled oscillators, numerical tests show that a simple iterative scheme consistently yields a stable and unique solution, 
provided that reasonable initial conditions are chosen. Quantitatively, the resulting thermodynamic quantities -free energy, average energy, and specific heat- are found to be quite accurate, with relative errors typically within a few percent, even at low temperatures.

Extending this approach to more realistic systems, particularly those involving a significantly larger number of oscillators, 
is the main challenge for future work. 
Preliminary results indicate that convergence issues may arise in such cases, suggesting that more advanced numerical algorithms may 
be required to ensure stability. In addition, the computational cost associated with evaluating the covariance matrix, which involves
$N$-dimensional integrals, may become a limiting factor. This limitation could potentially be avoided by modifying the criterion 
used to fit the exact potential with a Gaussian distribution (see Eq.(\ref{replaceV})), 
thus circumventing the evaluation of the covariance matrix.

Finally, as already noted in the introduction, a particularly promising application of our formula is the computation of 
the rovibrational partition function of large (bio)molecules. Achieving this will require extending the present formalism 
to handle the more complex kinetic energy operators that arise when expressed in internal coordinates. 
Initial developments in this direction are encouraging and suggest that such a generalization is feasible.

\acknowledgments{I would like to thank Dr. A. Scemama for interesting discussions and valuable help with the simulations. 
I also thank the Centre National de la Recherche Scientifique (CNRS) for its continued support. This work used the HPC resources from CALMIP (Toulouse) 
under allocation 2025-18005. I acknowledge funding from the European Research Council (ERC) 
under the European Union's Horizon 2020 research and innovation programme (Grant agreement No.~863481).
The data that supports the findings of this study
are available from the author upon reasonable request.
}
\appendix

\section{Two useful formulas}
\label{appendix_A}
As a first formula, let us show that 
\be
\sum_{i=1}^n \frac{ \qty(\sum_{j}  O^{ij})^2}{\lambda_i + \tau^2 x} = \frac{n^2}{\beta \tau x}
\label{A0}
\ee
This formula is easy to derive by remarking that the matrix $A$, which is invariant by translation, 
admits as one of its eigenstates the constant normalized vector, say
$u_1=(\frac{1}{\sqrt{n}},...,\frac{1}{\sqrt{n}})$ with eigenvalue $\lambda_1=0$. As a consequence, we 
have
$$
\sum_{j=1}^n  O^{1j}=  \sum_{j=1}^n  u_{1j}= \sqrt{n}
$$
The columns of the matrix $O^{ij}$ being orthogonal we have
$$
\sum_{j=1}^n O^{ij}=0 \;\; {\rm for} \;i \ne 1.
$$
Using this relation, only the component $i=1$ of the sum in Eq.(\ref{A0}) is not vanishing, thus leading to the desired equality.

As a second useful formula, let us first show that the quantity ${\cal A}$ defined as
\be
{\cal A} = \sum_{k=1}^N \frac{b^2_{gk}}{\omega^2_{gk}} - 2 \sum_{\substack{1 \leq k < l \leq N}} K_{gkl} {\bar x}_k {\bar x}_l + \sum_{k=1}^N  \frac{ \qty( \sum_l {\bar K}_{gl} S_{kl})^2}{\Omega^2_{gk}}
\label{A1}
\ee
is equal to the simpler expression
$$
{\cal A}=\qty(b_g, M^{-1} b_g)
$$
where 
\be
M_{kl} = \omega^2_{gk} \delta_{kl} + K_{gkl}.
\label{A3}
\ee
Let us first evaluate the third term of the R.H.S of Eq.(\ref{A1}). Defining the quantity $X$ as follows
\be
X \equiv \sum_{k=1}^N \frac{ \qty(\sum_{l=1}^N {\bar K}_{gl} S^T_{lk})^2} { \Omega^2_{gk}}
\ee
and using Eq.(\ref{Kbar}) we have
$$
X =  \sum_{l,m,i,j} \qty( \sum_{k=1}^N \frac{ S^T_{lk} S^T_{mk}}{ \Omega^2_{gk}} ) K_{gli} K_{gmj} {\bar x}_i {\bar x}_j.
$$
Using the fact that
$$
\sum_{k=1}^N \frac{S^T_{lk} S^T_{mk}}{ \Omega^2_{gk}} = M^{-1}_{lm}
$$
and the expression of the matrix $M$, Eq.(\ref{A3}), we get
$$
X= \sum_{i,j} \qty(  M_{ij} - \delta_{ij} \omega^2_{gj} - \delta_{ij} \omega^2_{gi} +  M^{-1}_{ij} \omega^2_{gi}  \omega^2_{gj} ) {\bar x}_i {\bar x}_j
$$
Now, using the definition of ${\bar x}_i$, Eq.(\ref{xbar}), we get
$$
X= ({\bar x}, M {\bar x}) - 2 \sum_i \frac{ b^2_{gi}}{ \omega^2_{gi}} + \sum_{i,j} b_{gi} M^{-1}_{ij} b_{gj}.
$$
Putting this expression into expression (\ref{A1}) for ${\cal A}$ we obtain our final formula
$$
{\cal A}=\qty(b_g, M^{-1} b_g)
$$
\section{Path Integral Monte Carlo}
\label{appendix_B}
Path Integral Monte Carlo (PIMC) is a standard numerical method for computing quantum thermodynamic properties; for a detailed presentation, 
see {\it e.g.} \cite{Ceperley1995}. In our implementation we use a Fourier representation for the paths, a way of generating preferentially 
paths with large
scale deformations and, thus, an improved convergence for the estimators\cite{feynman1972statistical},\cite{Doll_1984},\cite{Doll_Freeman}. 
Using the notations of this work, Fourier modes are realized here 
by drawing independently the
transformed variables ${\tilde x}^i_k = \sum_{j=1}^n  O^{ij} x^j_k$ resulting from the diagonalization of the kinetic matrix $A$, 
Eq.(\ref{matrixA}). The probability density function is given by
\be
\pi({\tilde x}^i_k) \sim e^{-\frac{\lambda_i}{\tau} {\tilde x}^{2i}_k}.
\ee
In PIMC various estimators can be employed for computing the average energy and specific heat. Here, we use the following low-variance estimators.
For the average energy $E$ 
we use the so-called centroid virial estimator given by\cite{Shiga_2005},\cite{Moustafa_2024}
\be
E =  \langle \epsilon_{CV} \rangle = \frac{N}{2\beta} + \frac{1}{2} \langle {\bar W}_1 \rangle  +  \langle {\bar V} \rangle
\ee
where
\be
{\bar V}= \frac{1}{n} \sum_{i=1}^n V\qty({\bf x}^i)
\ee
and
\be
{\bar W}_1= \frac{1}{n} \sum_{i=1}^n \sum_{k=1}^N  \qty(x^i_k- {\bar x}_k) \frac{ \partial V}{\partial x_k}\qty({\bf x}^i)
\ee
with
\be
{\bar x}_k=  \frac{1}{n} \sum_{i=1}^n x^i_k.
\ee
For the specific heat, we use the so-called double centroid virial estimator given by\cite{Shiga_2005},\cite{Moustafa_2024}
\be
c_v= \beta^2 \qty[\langle \epsilon^2_{CV}\rangle -\langle \epsilon_{CV} \rangle^2 
- \frac{1}{4\beta} \langle 3 {\bar W}_1 + {\bar W}_2 \rangle ]
\ee
with
\be
{\bar W}_2= \frac{1}{n} \sum_{i=1}^n \sum_{k,l=1}^N \qty(x^i_k- {\bar x}_k) \qty(x^i_l- {\bar x}_l)
\frac{ \partial^2 V}{\partial x_k \partial x_l}\qty({\bf x}^i)
\ee


\end{document}